\documentclass[preprint,11pt]{imsart}

\usepackage{amsthm,amsmath,natbib}
\usepackage{amsfonts}
\usepackage{amssymb}

\RequirePackage[colorlinks,citecolor=blue,urlcolor=blue]{hyperref}

\usepackage{tikz-cd}

\usepackage[a4paper,includeheadfoot,margin=2.54cm]{geometry}

\startlocaldefs
\newcommand{\vekk}[1]{}
\usepackage{amsthm}

 % ... 

\newcommand{\abs}[1]{\left| #1 \right|} 
\newcommand{\RealN}{{\mbox{$\mathbb R$}}} % Real numbers
\newcommand{\st}{\operatornamewithlimits{{\mid}}}
\newcommand{\Normal}{\mathop{\text{N}}} % Normal, or Gaussian
\newcommand{\nc}{\newcommand}
\nc{\beqns}{\begin{eqnarray*}}
\nc{\eeqns}{\end{eqnarray*}}
\nc{\beqn}{\begin{eqnarray}}
\nc{\eeqn}{\end{eqnarray}}
\nc{\beq}{\begin{equation}}
\nc{\eeq}{\end{equation}}

\newcommand{\bes}{ \begin{equation} \begin{split} } %Nummerert slutt.
\newcommand{\ees}{ \end{split} \end{equation} } %Nummerert slutt.

\newcommand{\be}[1]{\begin{equation}\label{eq#1}} 
\newcommand{\ee}{\end{equation}}%Nummerert slutt.

\theoremstyle{plain}

\theoremstyle{remark}

\endlocaldefs

\begin{document}

\begin{frontmatter}

% "Title of the Paper"
%\title{Improper priors and conditional expectation}
%\title{Conditional expectation, sufficiency, and improper priors}
\title{Fiducial on a string\\
{\small Revisiting Fisher's fiducial inference}}

\runtitle{Fiducial on a string}

\begin{aug}
% indicate corresponding author with \corref{}
\author{\fnms{Gunnar} \snm{Taraldsen}\corref{}\ead[label=e1]{Gunnar.Taraldsen@ntnu.no}}
\and
%\address[a]{\printead{e1};\printead{e2}}
\author{\fnms{Bo Henry} \snm{Lindqvist} \ead[label=e2]{bo@math.ntnu.no}}
%\and
%\author{\fnms{???} \snm{???}\thanksref{b}\ead[label=e2]{???}}
%\address[a]{\printead{e1}}
%\address[b]{\printead{e2}}

\address{Trondheim, Norway.
\printead{e1,e2}}

\runauthor{Taraldsen and Lindqvist}

\affiliation{Norwegian University of Science and Technology}

\end{aug}

\begin{abstract}
The fiducial argument of \citet{FISHER}
has been described as his biggest blunder,
but the recent review of \cite{HannigIyerLaiLee16review}
demonstrates the current and increasing interest in this brilliant idea. 
This short note analyses an example introduced
by \citet{Seidenfeld92fiducial} where the 
fiducial distribution is restricted to a string.

\vekk{
The abstract should summarize the contents of the paper.
It should be clear, descriptive, self-explanatory and not longer
than 200 words. It should also be suitable for publication in
abstracting services. Please avoid using math formulas as much as possible.

This is a sample input file.  Comparing it with the output it
generates can show you how to produce a simple document of
your own.
}

\end{abstract}


%\begin{abstract}
%\end{abstract}

%\begin{keyword}
%\kwd{}
%\kwd{}
%\end{keyword}

% history:
% \received{\smonth{1} \syear{0000}}

\tableofcontents

\end{frontmatter}

\section{The problem}
\label{sProb}

\begin{figure}[h]
\begin{center}
\includegraphics[width=0.4\textwidth]{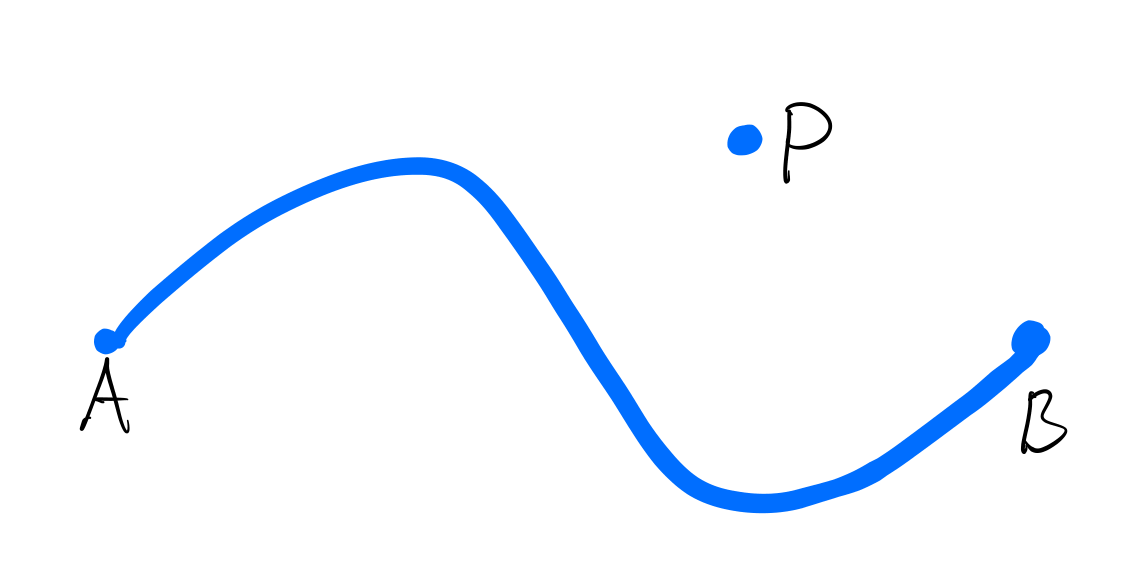} 
\caption{
% A fiducial estimate $P$ and a string containing the true parameter.
{\bf How can a fiducial distribution on the string be derived?}
}
\label{fig1}
\end{center}
\end{figure}

Assume that a fiducial distribution located at $P$ has been derived,
but that it is known that the parameter lies on a string connecting points $A$
and $B$ as illustrated in Figure~\ref{fig1}.

\citet[p.138-142]{FISHER} considers this problem for the particular case
where the initial fiducial is bivariate normal with mean $P$.
For the case of a straight line and a circle he derives a fiducial
from sufficiency and ancilarity respectively.
For the general case \citet[p.142]{FISHER} indicates that 
the fiducial can be calculated from the likelihood for each point on the curve.

\citet[Example 5.1]{Seidenfeld92fiducial} considers the case where
the observations are given by
\be{1}
x = \mu (t) + u, \;\; t \in \RealN
\ee
where $\mu_1 (t) = t^3$, $\mu_2 (t) = t$, 
and $u$ is drawn from $\Normal (0, I)$. 
Using Bayes' theorem argumentation he arrives at two
contradicting fiducial distributions.
The first corresponds to the Bayes posterior from 
a uniform prior law for $t^3$, 
and the other corresponds to a uniform prior law for $t$.
\citet[p.367]{Seidenfeld92fiducial} notes that this
example is challenging for a wide variety of
what Savage called ``necessitarian'' theories:
Theories that try to find privileged distributions to represent ``ignorance''.

\section{Conditional fiducial inference} 
\label{sCondFid}

\begin{figure}[h]
\begin{center}
\includegraphics[width=0.7\textwidth]{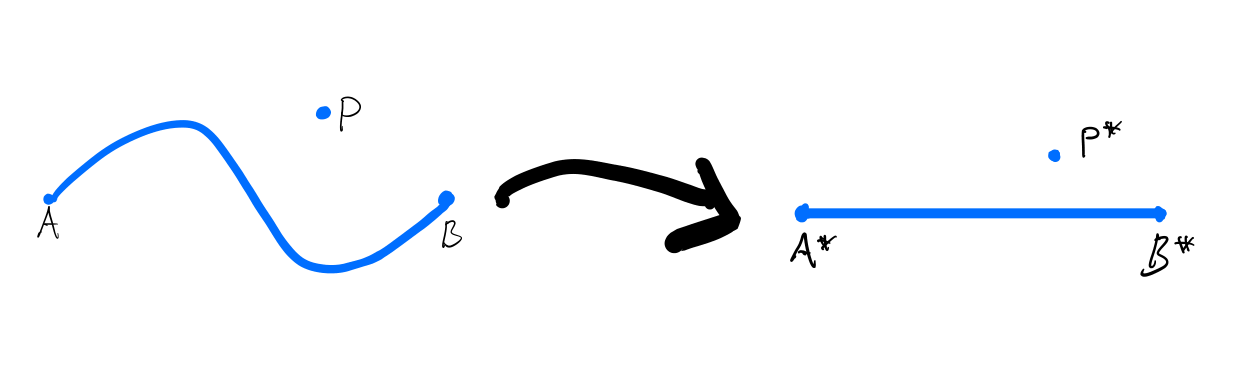} 
\caption{
% A fiducial estimate $P$ and a string containing the true parameter.
{\bf A transformation of the string problem.}
}
\label{fig2}
\end{center}
\end{figure}

The curve in Figure~\ref{fig1} can be transformed to a line segment
as indicated in Figure~\ref{fig2}.

The problem for the line segment can be solved by introducing coordinates
$(\theta_1^*, \theta_2^*)$ such that the line segment is given by
$\theta_2^* = 0$ and $A^* \le \theta_1^* \le B^*$.
A fiducial on a string is then determined by the conditional law
\be{2}
\Theta \st (\Theta_2^* = 0, A^* \le \Theta_1^* \le B^*)
\ee 
If $\theta^* \mapsto \theta$ has a derivative $\theta'$ and
the initial fiducial is given by a probability density $\pi (\theta)$, 
then the conditional law will have a density proportional to $\pi^* (\theta_1^*, 0)$
where 
\be{3}
\pi^* (\theta^*) = \pi (\theta) \abs{\theta'}
\ee
The initial fiducial from equation~(\ref{eq1}) is given by
%
%\be{3}
$\theta = x - u$
%\ee
%
where $u$ is sampled from the bivariate normal density $f$.
The initial density of $\theta$ is then
%
%\be{4}
$\pi (\theta) = f( x - \theta)$.
%\ee
%
A particularly simple transformation is given on the form
$\theta = \mu (t) + \nu (s)$ with $\nu (0) = 0$,
$\theta_1^* = t$, and $\theta_2^* = s$.
The resulting density for $t$ is then proportional to
\be{5}
f (x - \mu (t)) \abs{\dot{\mu}_\perp}
\ee
The last factor is the component of $\dot{\mu} (t)$
orthogonal to $\dot{\nu}(0)$.
It is proportional to
$\abs{\dot{\mu}_1 \dot{\nu}_2 - \dot{\mu}_2 \dot{\nu}_1 }$
evaluated at $s=0$. 
The two alternative fiducial distributions obtained by \citet{Seidenfeld92fiducial} are given by
$\nu$ corresponding to shifts in the two coordinate directions.
Generally, restricted to the simple transformation form, 
the last factor is proportional to any non-zero linear combination of the coordinates of $\dot{\mu}$.  
A more general solution is given by equation~(\ref{eq3}).

\section{A fiducial argument}
\label{sFidArg}

Assume that an initial fiducial model is given by a 
quasi-group multiplication \citep{TaraldsenLindqvist13fidopt}
\be{6}
x = \theta u
\ee
In this case,
the fiducial distribution of $u$ is obtained by making
the judgment that it equals its original sampling distribution.
The initial fiducial distribution of $\theta$ is then uniquely 
determined by equation~(\ref{eq6}) when $x$ is the fixed observed value and $u$ is from 
its original sampling distribution.   

Consider now as previously that $\theta$ is restricted to values
with $\theta = \mu (t)$ for an underlying parameter $t$.
This gives, for a fixed observed $x$, 
a corresponding parameterization of possible values of $u$ determined by
equation~(\ref{eq6}).
Introduce now a transformation $u \mapsto u^*$ so that
the restriction on $u$ corresponds to $u_2^* = 0$ and $a^* \le u_1^* \le b^*$.
The resulting fiducial distribution for $u$ after observing $x$ is then the conditional distribution
\be{7}
U \st (U_2^* = 0, a^* \le U_1^* \le b^*)
\ee

If $u^* \mapsto u$ has a derivative $u'$ and
the initial fiducial is given by a probability density $f (u)$, 
then the conditional law will have a density proportional to $f^* (u_1^*, 0)$,
where 
%
%\be{3}
\be{8}
f^* (u^*) = f (u) \abs{u'}
\ee
This is in complete analogy with equation~(\ref{eq3}), 
and leads in particular to the two fiducial distributions of 
\citet{Seidenfeld92fiducial} as before
by consideration of a transformation on the form
$u = x - (\mu (t) + \nu (s))$ for this additive group case.
The transformation $u^* \mapsto u$ depends in this case explicitly on $x$.

The results so far correspond to $\theta = \mu (t)$ and 
a formal Bayes prior density for $t$ on the form
\be{9}
h (\dot{\mu} (t)) = \abs{c_1 \dot{\mu}_1 (t) + c_2 \dot{\mu}_2 (t)}
\ee
Equation~(\ref{eq5}) corresponds to a case where $c$ does not depend on $t$,
but equation~(\ref{eq3}) corresponds to a case with $t$ dependence of $c$.
Equation~(\ref{eq8}) can lead to cases where the weight $c$ 
also depends on $x$: A data dependent prior.
This is as described more generally by \citet{HannigIyerLaiLee16review} using a different,
but related approach.

The Jeffreys prior is on the form
\be{10}
h (\dot{\mu}) = \abs{\dot{\mu}}
\ee
The fiducial resulting from the arguments leading to equation~(\ref{eq9})
will always lead to a proper posterior as dictated by equations~(\ref{eq6}-\ref{eq7}).
If equation~(\ref{eq6}) corresponds to a locally compact group, 
then the initial fiducial is a Bayes posterior and the final fiducial resulting from 
equation~(\ref{eq9}) is then also a Bayes posterior.
 
\section{Discussion and conclusion}
\label{sDisc}

This note was initiated due to comments from Teddy Seidenfeld during
the 4th Bayesian, Fiducial and Frequentist (BFF4) workshop 
at Harvard University in May 2017.
We gave an invited talk where the concept of a conditional fiducial model
was presented: A fiducial model together with a condition $C = c$
\citep{TaraldsenLindqvist17confid}.
The fiducial is not uniquely given by the additional demand that the parameter
is restricted to be on a given curve.
It is, however, uniquely given if this is reformulated by a condition $C=c$.  
In the setting given by this note, it follows that it corresponds
to a formal prior on the form
\be{11}
h = \abs{\nabla C} \abs{\dot{\mu}}
\ee
Note that $\nabla C$ is a normal vector to the curve tangent
vector $\dot{\mu}$,
and that it is possible to choose a path-length parameterisation so that $\abs{\dot{\mu}} = 1$.
A condition $C$ with $\abs{\nabla C}$ constant on the curve gives the Jeffreys prior in
equation~(\ref{eq10}),
and other choices give the general result corresponding to equation~(\ref{eq3}).

We hereby acknowledge most fruitful discussions with Teddy Seidenfeld, 
Jan Hannig, and the other participants at the Harvard BFF4 workshop. 

\bibliography{bib,gtaralds}

\begin{thebibliography}{5}
% BibTex style file: imsart-nameyear.bst, 2013-01-28
% Default style options (sort=1,type=nameyear).
% Used options (sort=1,type=nameyear).

\bibitem[\protect\citeauthoryear{Fisher}{1973}]{FISHER}
\begin{bbook}[author]
\bauthor{\bsnm{Fisher},~\bfnm{R.~A.}\binits{R.~A.}}
(\byear{1973}).
\btitle{{Statistical methods and scientific inference}}.
\bpublisher{Hafner press}.
\end{bbook}
\endbibitem

\bibitem[\protect\citeauthoryear{Hannig
  et~al.}{2016}]{HannigIyerLaiLee16review}
\begin{barticle}[author]
\bauthor{\bsnm{Hannig},~\bfnm{J.}\binits{J.}},
  \bauthor{\bsnm{Iyer},~\bfnm{H.}\binits{H.}},
  \bauthor{\bsnm{Lai},~\bfnm{R.~C.~S.}\binits{R.~C.~S.}} \AND
  \bauthor{\bsnm{Lee},~\bfnm{T.~C.~M.}\binits{T.~C.~M.}}
(\byear{2016}).
\btitle{Generalized Fiducial Inference: A Review and New Results}.
\bjournal{Journal of the American Statistical Association}
\bvolume{111}
\bpages{1346-1361}.
\end{barticle}
\endbibitem

\bibitem[\protect\citeauthoryear{Seidenfeld}{1992}]{Seidenfeld92fiducial}
\begin{barticle}[author]
\bauthor{\bsnm{Seidenfeld},~\bfnm{Teddy}\binits{T.}}
(\byear{1992}).
\btitle{{R. A. Fisher's Fiducial Argument and Bayes' Theorem}}.
\bjournal{Statistical Science}
\bvolume{7}.
\bdoi{10.2307/2246072}
\end{barticle}
\endbibitem

\bibitem[\protect\citeauthoryear{Taraldsen and
  Lindqvist}{2013}]{TaraldsenLindqvist13fidopt}
\begin{barticle}[author]
\bauthor{\bsnm{Taraldsen},~\bfnm{G.}\binits{G.}} \AND
  \bauthor{\bsnm{Lindqvist},~\bfnm{B.~H.}\binits{B.~H.}}
(\byear{2013}).
\btitle{{Fiducial theory and optimal inference}}.
\bjournal{Annals of Statistics}
\bvolume{41}
\bpages{323--341}.
\end{barticle}
\endbibitem

\bibitem[\protect\citeauthoryear{Taraldsen and
  Lindqvist}{2017}]{TaraldsenLindqvist17confid}
\begin{barticle}[author]
\bauthor{\bsnm{Taraldsen},~\bfnm{G.}\binits{G.}} \AND
  \bauthor{\bsnm{Lindqvist},~\bfnm{B.~H.}\binits{B.~H.}}
(\byear{2017}).
\btitle{Conditional fiducial models}.
\bjournal{Journal of Statistical Planning and Inference}
\bvolume{(accepted)}.
\end{barticle}
\endbibitem

\end{thebibliography}
%\bibliographystyle{plainnat} 
%\bibliographystyle{asa}
%\bibliography{bib}
%\bibliographystyle{chicago} 
\bibliographystyle{imsart-nameyear}

\end{document}